\documentclass[useAMS,usenatbib]{mn2e}

\usepackage{txfonts}
\usepackage[dvips]{graphicx}
\usepackage{epsfig}
\usepackage{natbib}
\usepackage{subfigure}
\usepackage{multirow}
\bibpunct{(}{)}{;}{a}{}{,}

\newcommand{\sref}[1]{Section \ref{#1}}

\newcommand{\apj}{ApJ}
\newcommand{\apjl}{ApJL}
\newcommand{\apjs}{ApJS}
\newcommand{\aj}{AJ}
\newcommand{\mnras}{MNRAS}
\newcommand{\nat}{Nature}

\newcommand{\aap}{A\&A}
\newcommand{\apss}{Ap\&SS}
\newcommand{\aaps}{A\&AS}

\title[Galaxy clustering]%
{The influence of superstructures on bright galaxy environments: clustering properties}

\author[Yaryura et al.]
  {C.Y.~Yaryura$^{1}$, M.~Lares$^{1}$, H.E. Luparello$^{1}$,D.J.~Paz$^{1}$, 
  D.~G.~Lambas$^{1}$, N.~Padilla$^{2,3}$ and M.A.~Sgr\'o$^{1}$\\
 $^{1}$Instituto de Astronom\'{\i}a Te\'{o}rica y Experimental
  (CONICET-UNC). Observatorio Astron\'{o}mico de C\'{o}rdoba, 
  Laprida 854, X5000BGR, C\'{o}rdoba, Argentina\\
 $^{2}$Departamento de Astronom\'{\i}a y Astrof\'{\i}sica, Pontificia
 Universidad Cat\'olica de Chile, Santiago, Chile\\
 $^{3}$Centro de Astro-Ingenier\'{\i}a, Pontificia Universidad Cat\'olica 
 de Chile, Santiago, Chile }

\date{Released 2002 Xxxxx XX}
\pagerange{\pageref{firstpage}--\pageref{lastpage}} \pubyear{2002}

\def\LaTeX{L\kern-.36em\raise.3ex\hbox{a}\kern-.15em
    T\kern-.1667em\lower.7ex\hbox{E}\kern-.125emX}

\begin{document}
\label{firstpage}
\maketitle

\begin{abstract}
We analyse the dependence of clustering properties of galaxies as a function of
their large--scale environment.
In order to characterize the environment on large scales, we use the catalogue
of future virialized superstructures (FVS) by Luparello et al. and
separate samples of luminous galaxies according to whether or not they belong
to FVS.
In order to avoid biases in the selection of galaxies, we have constructed
different subsamples so that the distributions of luminosities and masses are
comparable outside and within FVS.
As expected, at large scales, there is a strong difference between the
clustering of galaxies inside and outside FVS. 
However, this behaviour changes at scales \mbox{r $\le$ 1 $h^{-1}$ Mpc}, where
the correlations have similar amplitudes.
The amplitude of the two--halo term of the correlation function for objects
inside FVS does not depend on their mass, but rather on that of the FVS. 
This is confirmed by comparing this amplitude with that expected from extended
Press-Schechter fits.
In order to compare these observational results with current models for
structure formation, we have performed a similar analysis using a
semi--analytic implementation in a \mbox{$\Lambda$CDM} cosmological model. 
We find that the cross--correlation functions from the mock catalogue depend on
the large--scale structures in a similar way to the observations.

From our analysis, we conclude that the clustering of galaxies within the
typical virialized regions of groups, mainly depends on the halo mass,
irrespective of the large--scale environment.
\end{abstract}


%····································································

%····································································
%                                00SECTIONS
\section{Introduction} \label{S_intro}
%{{{S_intro*/

The large--scale structure of the Universe, and in particular the largest
virialized systems, are a key probe of the evolution of the density field from
the primordial fluctuations, and consequently of the nature of the energy
content of the Universe and the long range action of gravity.
Thanks to the widely available galaxy redshift surveys performed in the last
years, as Las Campanas Redshift Survey \citep{Shectman:1996}, the 2-degree
Field Galaxy Redshift Survey \cite[2dFGRS,][]{Colless:2001} and the Sloan
Digital Sky Survey \cite[SDSS,][]{York:2000, Stoughton:2002}, we know that the
large--scale structure of the Universe appears as a network made up of walls,
filaments, knots and voids \citep{Joeveer:1978, Gregory:1978, Zeldovich:1982,
deLapparent:1986}.
The nodes are the intersections of walls and filaments, so they are the highest
density regions.
It is these regions, or some part of them, which we call superstructures.
In a \mbox{$\Lambda$CDM} cosmological model, where the dynamics of the Universe
is expected to be dominated by the accelerated expansion, superstructures can
be defined as the currently overdense regions that will be bound and virialized
systems in the future \citep{Busha:2005, Dunner:2006, Araya-Melo:2009,
Luparello:2011}. 
These superstructures provide important information about the matter
distribution on cosmological scales, allowing precise analysis of the
cosmological model \citep{Kolokotronis:2002, Einasto:2007b, Araya-Melo:2009,
Einasto:2011, Sheth:2011}. 
The clustering properties of galaxies on scales smaller than the size of
superstructures are key to observationally constrain the accretion processes
that give rise to luminous galaxies  \citep{Springel:2005, Bildfell:2011,
Fontanot:2009}.
In this context, the local environment is claimed to be the principal 
factor in defining the global properties of galaxies and their distribution
\citep{Grutzbauch:2011, Kimm:2009, Blanton:2005, Parker:2007, Park:2009}.

In fact, several models of galaxy formation assume that galaxy properties are
determined by the haloes where they formed, and not by the large
scale environment that surrounds them \citep{Berlind:2003, Yang:2003, 
Baugh:2006, Kauffmann:1997}.
In this context, the population of galaxies in a halo of a given mass
should be independent of the location of the halo.
However, in the last years there were different studies, both
observational and using simulations, which show that the galaxy
properties, such as the star formation rate or galaxy colours, depend
on large--scale structure \citep{Park:2009, Binggeli:1982, Donoso:2006, 
Crain:2009, White:2010}.

The formation and evolution of systems that are embedded in 
superstructures could be 
conditioned by these large overdensities \citep[][ and references 
therein]{Hoffman:2007, Araya-Melo:2009, Bond:2010}.
Galaxies within them also show different properties and 
spatial distributions \citep{Einasto:2003c, Wolf:2005, Haines:2006,
Einasto:2007c, Porter:2008, Tempel:2009, Fleenor:2010,
Tempel:2011, Einasto:2011}.

Recent galaxy redshift surveys \citep{York:2000, Colless:2001} sample
a sufficiently large volume to allow the study of galaxy
formation and evolution from a statistical perspective and unveil its
details with increasing confidence.
The aim of this work is to use SDSS--DR7 catalogue \citep{Abazajian:2009} 
to study the influence of superstructures on galaxies.
We will study the dependence of the clustering of faint galaxies (tracers,
\mbox{$-21.0<M_r<-20.5$}) around brighter galaxies (centres,
\mbox{$-23.0<M_r<-21.0$}) according to whether the centre galaxies are part
of superstructures.
In this work, we will analyse the clustering properties of galaxies
located in large structures which are still undergoing a 
virialization process.
Then we can study the differences in the small scale clustering of 
these galaxies compared to galaxies in the field.
For this study, we use the superstructures identified by
\cite{Luparello:2011}, dubbed Future Virialized Structures (FVS), 
identified in the SDSS--DR7 and \cite{Zapata:2009} groups which 
will be used to characterize the virial mass of the groups 
which host our galaxies.

This paper is organized as follows.
In \sref{S_data} we describe data collected for
groups of galaxies and future virialized structures.
We then analyse the clustering properties using the cross-correlation 
function, as described in \sref{S_method}, and show results in 
\sref{S_results}.
We summarize and discuss the results in \sref{s_conclusus}.

%}}}S_intro*/

\section{Data and Samples} \label{S_data}
%{{{S_data*/

\subsection{SDSS--DR7 Galaxy Catalogue} \label{SSDS-DR7} 

The Sloan Digital Sky Survey \citep{York:2000} is one of the largest and most
ambitious surveys carried out so far. 
It has deep multi--colour images covering more than one quarter of the sky, and
spectra for about 930000 galaxies and 120000 quasars.
It is one of the largest data sets produced and contains images, image
catalogues, spectra and redshifts. 
In this work we use the Seventh Data Release \cite[DR7,][]{Abazajian:2009},
which is publicly available \footnote{http://www.sdss.org/dr7}.

The spectroscopic galaxy catalogue comprises a footprint area of 9380 sq.deg.
The limiting apparent magnitude for the spectroscopic catalogue in the r--band
is 17.77 \citep{Strauss:2002}.
We use a more conservative limit of $17.5$ to ensure completeness. 
Also, in order to avoid saturation effects in the photometric pipeline, we
consider galaxies fainter than \mbox{r = 14.5}.
The image deblending software often fragments images of bright galaxies with
substructure, so our cuts prevent possible artifacts in the final catalogue
\footnote{http://www.sdss.org/dr7/products/general/target\_quality.html}.

\subsection{SDSS--DR7 Superstructures} \label{superstructures}

The maps of the Universe depict a complex network of filaments and
voids \citep[e.g.][]{Einasto:1996, Colless:2001, Jaaniste:2004, 
Einasto:2006, Abazajian:2009}, where clusters of galaxies are 
preferentially located at the nodes of filamentary structures
\citep{Gonzalez:2010, Murphy:2011}.
This picture was possible only after extended galaxy surveys 
were completed, and it is also 
supported by numerical experiments 
consistent with the standard cosmological model
\citep{Frisch:1995, Bond:1996, Seth:2003, Shandarin:2004}.

Large scale structures, which we will refer to as superstructures, 
have been studied with a variety of methods.
The first statistical studies of superstructures were performed by
linking Abell cluster positions \citep{Zucca:1993, Einasto:1997}. 
Later, the realization of wide-area surveys of galaxies with redshift
as Las Campanas Redshift Survey \citep{Shectman:1996}, the 2-degree
Field Galaxy Redshift Survey \citep[2dFGRS,][]{Colless:2001} and the
Sloan Digital Sky Survey \citep[SDSS,][]{York:2000, Stoughton:2002}, allowed 
the identification of superstructures directly from the large-scale
galaxy distribution. 
\cite{Einasto:2007} identified superclusters in the 2dFGRS using a
density field method, and \cite{CostaDuarte:2010} studied the
morphology of superclusters of galaxies in the SDSS. 
The largest catalogue of superclusters has been constructed by
\cite{Liivamagi:2010}, who implemented the density field method on the
SDSS--DR7 main and luminous red galaxy samples.
In all cases, superclusters are operationally defined as objects
within a region of positive galaxy density contrast, and thus are
subject to a certain degree of arbitrariness in their detection.

Within the \mbox{$\Lambda$CDM} Concordance Cosmological Model, an accelerated
expansion dominates the present and future dynamics of the universe and thus
determines the nature of gravitationally bound structures.
Therefore, an alternative definition of these large-scale structures is that of
overdense regions in the present-day universe that will become bound and
virialized structures in the future.
Thus, under the assumption that the luminosity is a somewhat unbiased tracer of
mass on large scales, the integrated luminosity density of galaxies is commonly
used as an indicator of mass density.
\cite{Luparello:2011} applies the luminosity density field method to identify
FVS in large spectroscopic galaxy surveys. 
The identification procedure is based on the theoretical criteria of
\cite{Dunner:2006}. 
Using numerical simulations, they establish the minimum mass overdensity
neccesary for a structure to remain bound in the future. 
\cite{Luparello:2011} use this mass overdensity criteria to calibrate the
luminosity--density threshold, and identify superstructures as systems that are
likely to evolve into virialized structures in the distant future, in the
assumed cosmology.
The catalogue of FVS was extracted from a volume-limited sample of galaxies
from the SDSS--DR7, in the redshift range \mbox{$0.04<z<0.12$}, with a limiting
absolute magnitude of \mbox{$M_r <-20.47$}. 
The luminosity-density field is constructed on \mbox{1 h$^{-1}$ Mpc} cubic
cells grid, applying an Epanechnikov kernel of \mbox{ $r_0$ = 8 h$^{-1}$ Mpc}
(equation 3 of \cite{Luparello:2011}). 
The luminosity overdensity threshold is fixed at
\mbox{$D_T=\rho_{lum}/\bar{\rho}_{lum}=5.5$}, and the structures are
constructed by linking overdense cells with a simple version of a ''Friends of
Friends'' algorithm.
In order to avoid contamination from smaller systems, they also assign a lower
limit for the total luminosity of a structure at \mbox{L$_{struct}\geq$
10$^{12}$L$_{\odot}$}. 
The main catalogue of superstructures has completness over 90 per cent and
contamination below 5 per cent, according to calibrations made using mock
catalogues.
The volume covered by the catalogue is \mbox{3.17 $\times$ $10^7$ (h$^{-1}$
Mpc)$^3$}, within which 150 superstructures were identified, composed by a
total of 11394 galaxies.
FVS luminosites vary between \mbox{ $10^{12}$ L$_{\odot}$} and \mbox{$\simeq$
10$^{14}$ L$_{\odot}$}, and their volumes range between \mbox{10$^2$ (h$^{-1}$
Mpc)$^3$} and  \mbox{10$^5$ (h$^{-1}$ Mpc)$^3$}.
Because of the luminosity density field dependence, \cite{Luparello:2011}
analysed 3 samples with different luminosity thresholds, dubbed S1, S2 and S3,
described in \mbox{table 1} of their paper. 
We will consider sample S2 in our analysis, which contains 89513 galaxies with
\mbox{$M_r < -20.47$} in the redshift range \mbox{$0.04 < z < 0.12$}.

\subsection{SDSS--DR7 Galaxy Groups} \label{galaxy_groups} 
 
The aim of this paper is to analyse the clustering of galaxies
considering different samples.
Some of these samples are selected taking into account the mass of the
systems considered. 
To estimate the mass distributions of the samples
we use the estimated mass of galaxy groups.
The galaxy groups correspond to those identified in the SDSS galaxy
catalogue presented by \cite{Zapata:2009}, extended to cover the 
SDSS--DR7.
\cite{Zapata:2009} identified groups using the same method as
\cite{Merchan&Zandivarez:2005}, with a ''Friends of Friends'' 
algorithm, with a variable projected linking length and a fixed 
radial linking length.
These were set by Merchan $\&$ Zandivarez to obtain a sample as
complete as possible and with low contamination (95 per cent  and 8 per cent,
respectively).
The catalogue contains 83784 groups with at least 4 members, and
is limited to redshift \mbox{$z < 0.12$}.

\subsection{Synthetic catalogue and simulation} \label{mock_data} 

In order to test the procedures implemented in this work, we constructed a
mock galaxy catalogue from a semi-analytic model of
galaxy formation \citep[GALFORM,][]{Bower:2008} applied to the Millennium 
simulation \citep{Springel:2005}, which adopts a \mbox{$\Lambda$CDM}
concordance cosmological model.
The semi-analytic model of galaxy formation collects information from
the merger trees extracted from the simulation, and generates a
population of galaxies within the simulation box.
The mock catalogue is set up to mimic the geometry of the SDSS--DR7
footprint and reproduces the dilution in the number of galaxies
with redshift.

%--------------------------------------------------------------------</T>
%}}}S_data*/

\section{Method} \label{S_method}
%{{{S_method*/

The clustering of galaxies is analysed using the
two--point correlation function.
Early studies of the spatial distribution of galaxies using this 
technique were presented by \cite{Peebles:1980}. 
The two--point correlation function, $\xi (r)$, is defined as the excess 
of the probability, $\delta P$, of finding a galaxy, defined as 
\textit{tracer}, at a given distance from another galaxy, 
dubbed the \textit{centre}.
Thus, 

\begin{equation}
\delta P = n_g (1+\xi(r)) dV,
\end{equation}

\noindent
where $dV$ is a volume element and $n_g$ is the mean number density of
tracer galaxies.
There are several estimators of this function \citep{Kerscher:2000} that compute 
this probability by counting object pairs. In this work, we use one of the most 
commonly used estimators to determine the 
cross--correlation function, as defined by \cite{Davis:1983}:

\begin{equation}
\xi = \frac{D_{1}D_{2}}{R_{1}R_{2}} \frac{N_{R_{1}} N_{R_2}}{N_{D_{1}} N_{D_2}}  - 1 ,
\label{eq:sigma_s1}
\end{equation}

\noindent
where $D_{i} D_{j}$ is the average number of object pairs and $R_{i}R_{j}$ 
is the average number of pairs in a random sample. 
$N_{D_{i}}$ and $N_{R_{i}}$ represent the number of objects in the 
data catalogue and in the random sample, respectively. 
The random sample is generated with the same geometry as the real 
catalogue using the same angular and radial selection functions. 
In particular, the advantage of this estimator is that if the sample 
$D_{1}$ has few objects, $R_{1}$ can be selected to be larger 
than $D_{1}$, and thus minimize the noise in the $\xi$ calculation. 
Although there are more accurate estimators, like the one defined 
by \cite{Landy:1993}, differences between both estimators are 
negligible due to the large volume size of the SDSS \citep[e.g.][]{Paz:2011}.

In galaxy redshift catalogues, where the distance is estimated using the
spectroscopic redshift which includes a peculiar velocity component,
the three dimensional distribution of galaxies is affected by
distortions in the line of sight.
To take this into account we estimate the correlation function 
\mbox{$\xi (\sigma, \pi)$}, as a function of the 
projected ($\sigma$) and line of sight ($\pi$) distances.
As these distortions in redshift space occur only in the radial 
direction, we implement the inversion presented by \citet{Saunders:1992} 
to obtain the spatial correlation function $\xi(r)$ from 
\mbox{$\xi(\sigma,\pi)$}.
We integrate along the line of sight to obtain the projected correlation 
function $\Xi(\sigma)$:

\begin{equation}
\Xi(\sigma) = 2 \int^{\infty}_{0} \xi(\sigma, \pi) d\pi = 2 \int^{\infty}_{0} \xi(\sqrt{\sigma^{2} + y^{2}}) dy .
\label{eq:xi_sigma_int}
\end{equation} 
\noindent
In this work we use $\sigma = 0.3$ and $\pi = 0.3$ Mpc as 
lower integration limits for this equation, and  $\sigma = 30$ and $\pi = 30$ Mpc as upper limits. 
We use diferent binning schemes for $\xi(\sigma, \pi)$ according to the 
size of each sample, being the number of bins in the range 13 to 19.

Then, we can directly estimate the real space correlation function 
(\mbox{$\xi(r)$}) by the inversion of $\Xi(\sigma)$ assuming 
a step function $\Xi(\sigma_i) = \Xi_{i}$ in bins centered in $\sigma_{i}$ 
and interpolating between $r = \sigma_{i}$ values 
\citep[equation 26 of][]{Saunders:1992}:

\begin{equation}
\xi(r) = - \frac{1}{\pi} \sum_{j \geq i} \frac{\Xi_{j + 1} - \Xi_{j}}{ \sigma_{j + 1} -\sigma{j}} ln \left( \frac{ \sigma_{j + 1} + \sqrt{\sigma_{j + 1}^{2} - \sigma_{i}^2}}{ \sigma_{j} + \sqrt{\sigma_{j}^{2} - \sigma_{i}^2}} \right) .
\label{eq:xi_sigma_i}
\end{equation}
\noindent
This equation is a simple and direct way to invert the projected
correlation function.

In addition, we also compute the correlation function in 
redshift space, $\xi (s)$, where 
\mbox{$s^{2} = x^{2} + y^{2} - 2xycos\theta$} is the redshift-space 
distance; and $x$ and $y$ are the line of 
sight distances to each object of the pair, 
with angular separation given by $\theta = \sigma /y$.
$\xi(s)$ contains information about the peculiar velocities of 
the galaxies in the sample. 
%

%}}}S_method/*

\section{Results} \label{S_results} 
%{{{S_results/*

%%% SAMPLE DESCRIPTION %%%-------------------------------------------<T>
\begin{table*}
\begin{center}
%\begin{tabular}{ | l | l | l | l | l | l |}
\begin{tabular}{ | l | c | c | c | c | c |}
  \hline
   & galaxies & galaxies in Groups & Groups & $N_{in}$ & $N_{out}$ \\
  \hline \hline

  $-23.0 < M_{r} < -21.0$ & $g$ & & & {5563} & {33306}\\
  $-23.0 < M_{r} < -21.0$ & & $g_{G}$ & & {2013} & {4603}\\
  $-23.0 < M_{r} < -21.0$, same luminosity dist.& $gL$ & & & {5102} & {30526} \\
  $-23.0 < M_{r} < -21.0$, same mass dist.&  & $g_{G}M$ & & {1765} & {3900} \\
  $-23.0 < M_{r} < -21.0$, same mass and luminosity dists. & & $g_{G}ML$ & & {1765} & {3316}\\
  $-23.0 < M_{r} < -21.0$, same mass and luminosity dists., in mock & & $g_{G}ML-mock$ & & {5550} & {8711} \\

  \hline

  $>8$ members and same mass distribution   & & & $G8M$ & {509} & {1829}\\  
  $>8$ members and same mass dist., in mock & & & $G8M-mock$ & {708} & {2145} \\  
  \hline
  \end{tabular}%

  \caption{Description of samples of galaxies or groups used as
  centres in the computation of $\xi(r)$.
  Each name indicates objects (galaxies or groups) 
  inside and outside FVS by the ''in'' and ''out'' sufixes, 
  respectively.
  When indicated, subsamples are selected so that the
  distributions of galaxy luminosities or group masses inside and
  outside FVS are comparable. The last two columns indidate the 
 number of objects in each centre sample. }

\label{table:description} 
\end{center}
\end{table*}
%--------------------------------------------------------------------</T>

We select galaxies which are members of the superstructures defined by
\cite{Luparello:2011} to construct the samples of galaxies in FVS, while we
select galaxies which are members of the SDSS--DR7 galaxy catalogue but
are not members of FVS for the samples of galaxies outside
superstructures.
To analyse the clustering properties of both samples, we measure the
cross--correlation function, using faint galaxies with luminosity 
\mbox{$-21.0 < M_{r} < -20.5$} as tracers.
In both cases we select galaxies with luminosity 
\mbox{$-23.0 < M_{r} < -21.0$} as centres.
The centre and tracer galaxies are above the luminosity limit for the
volume limited sample of the SDSS--DR7 galaxy catalogue.

In Section \ref{sample_g}, we analyse the correlation functions around
centre galaxies inside and outside FVS.
The difference in the clustering of faint galaxies around bright
galaxies is not only produced by the selection of galaxies according
to their FVS membership, but also is related to other processes that
occur on smaller scales.
In particular, the luminosity of galaxies is known to correlate with
the clustering amplitude of galaxies \citep{Alimi_1988,
Zehavi:2005, Swanson_2008, Wang:2011, Zehavi:2010, Ross_2011}.
There is also evidence of the dependence of clustering with dark matter 
host mass \citep{Zehavi:2012}.
Then, in order to disentangle and quantify the contribution of
luminosity and mass to the total correlation amplitude, we
will analyse the effect of equating luminosity and mass distributions 
on the correlation signal in Sections \ref{sample_gL} and \ref{sample_gGML}.
In order to perform this analysis, we will use different samples of
centre galaxies selected according to different properties.
In Table \ref{table:description} we give a brief summary of the 
samples (galaxies or groups) which are used as centres in the 
computation of correlation functions.
Each sample name is used for two samples: one for galaxies located in 
a FVS, which we denote with the ''in'' subindex, and the other for 
galaxies which are not part of any FVS, identified with the ''out'' subindex.
The number of objects in each sample is also given in Table \ref{table:description}.

In Section \ref{sample_g} we will analyse samples $g_{in}$ and
$g_{out}$, which contain bright galaxies ($-23.0 <
M_{r} < -21.0$) of the SDSS catalogue.
In Section \ref{sample_gL} we will analyse samples $gL$ which are
defined from the $g$ samples, but with the additional restriction that
their luminosity distributions are comparable.
In Section \ref{sample_gG} we introduce the SDSS--DR7 galaxy
groups in the analysis and we separate the groups located inside and 
outside FVS. 
Then, we define the samples $g_{G}$
selecting bright galaxies contained in these groups.

To analyse effects on the galaxy selection, in Section \ref{sample_gGM} we
define new samples from the group catalogue. 
We redefine the group samples of Section \ref{sample_gG} in order to have the
same group virial mass distributions. 
The samples denoted by $g_{G}M$ are obtained by selecting the bright galaxies contained
in these groups.
To further constrain our samples, in Section \ref{sample_gGML} we redefine the
samples $g_{G}ML$ by adding the condition that the bright galaxies located in
groups with the same mass distributions also have the same luminosity
distributions, i.e, the $g_{G}ML$ samples contain bright galaxies with the same
luminosity distributions located in groups with the same mass distributions.
In Section \ref{sample_G8M} we select galaxy groups with at least 8 members, and
then divide them according to whether they are located in FVS or not, and we make
their mass distributions similar by selectively limiting the sample.
In this case, to define the $G8M$ samples we select geometrical centres of
groups instead of the bright galaxies, so that we can use virial 
mass estimated for groups to uncover mass effects.
In Section \ref{mock} we define the same samples as $g_{G}ML$ and $G8M$ but
taken from a mock catalogue. 
We name these samples as $g_{G}ML-mock$ and $G8M-mock$
respectively, and use them to test the reproducibility of our results by 
current models for structure formation.

%+++++++++++++++++++++++++++++++++++++++++++++++++++++++++++++++++++++++
\subsection{Clustering of faint galaxies around bright galaxies} 
\label{sample_g}

\begin{figure} 
   \centering
   \includegraphics[width=0.5\textwidth]{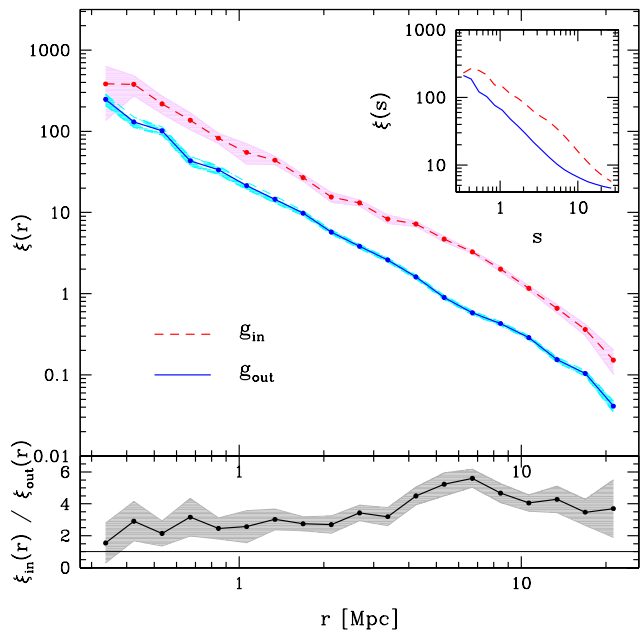} 
   \caption{
   Cross--correlation functions of SDSS--DR7 galaxies for the $g$ samples. 
   The dashed lines correspond to galaxies in sample $g_{in}$ 
   and the solid lines correspond to galaxies in sample $g_{out}$.
   }
   \label{fig:xi_sample_g} 
\end{figure}

\begin{figure*}
   \centering
   %\subfigure[Luminosity distributions for $g$ and $gL$ samples. ]{
   \subfigure[]{
   \label{fig:xi_sample_gL:a} 
   \includegraphics[width=0.48\textwidth]{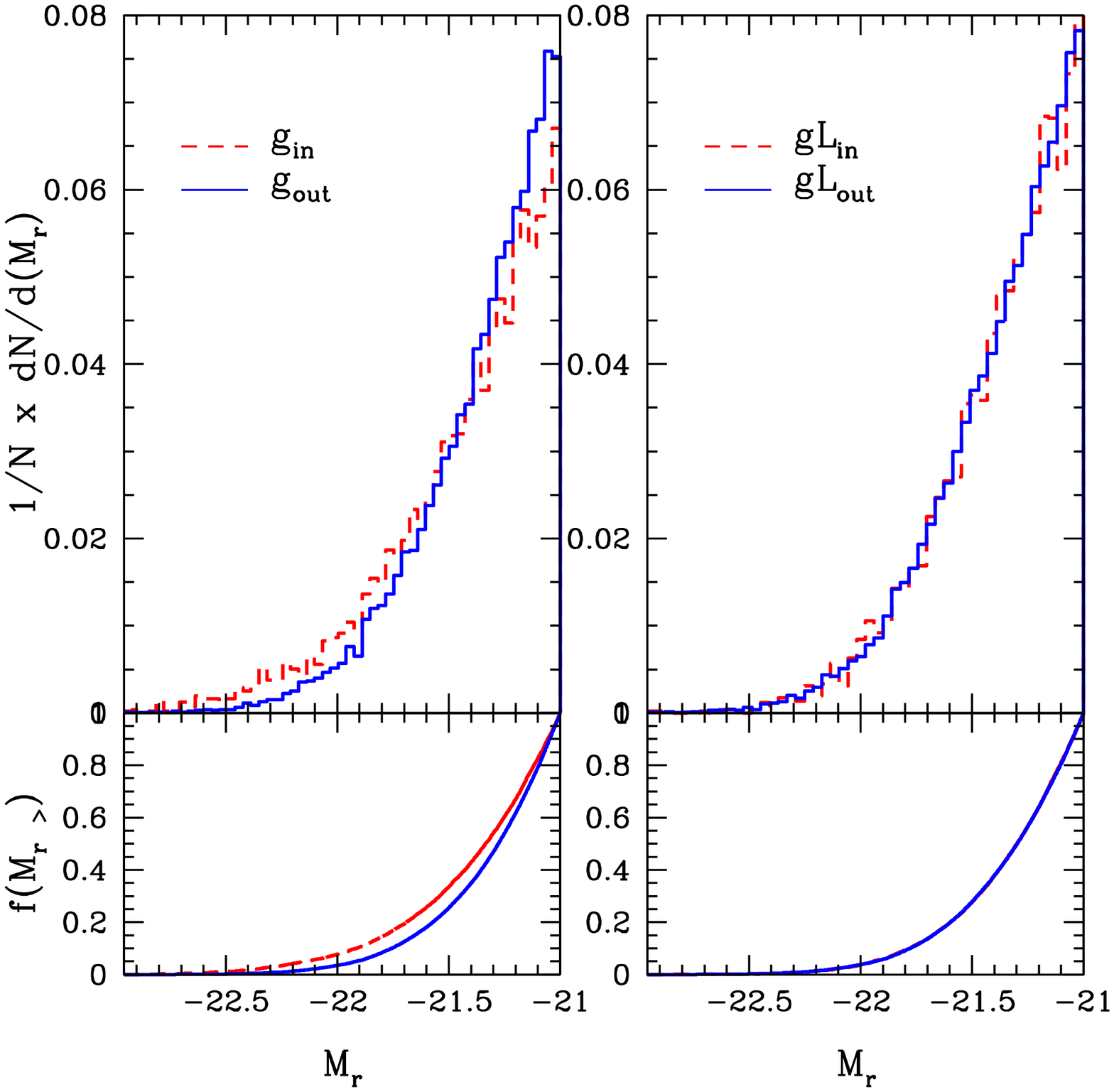}}
   \hspace{5pt}%
   %\subfigure[Cross correlation function for $gL$ samples.]{
   \subfigure[]{
   \label{fig:xi_sample_gL:b} 
   \includegraphics[width=0.48\textwidth]{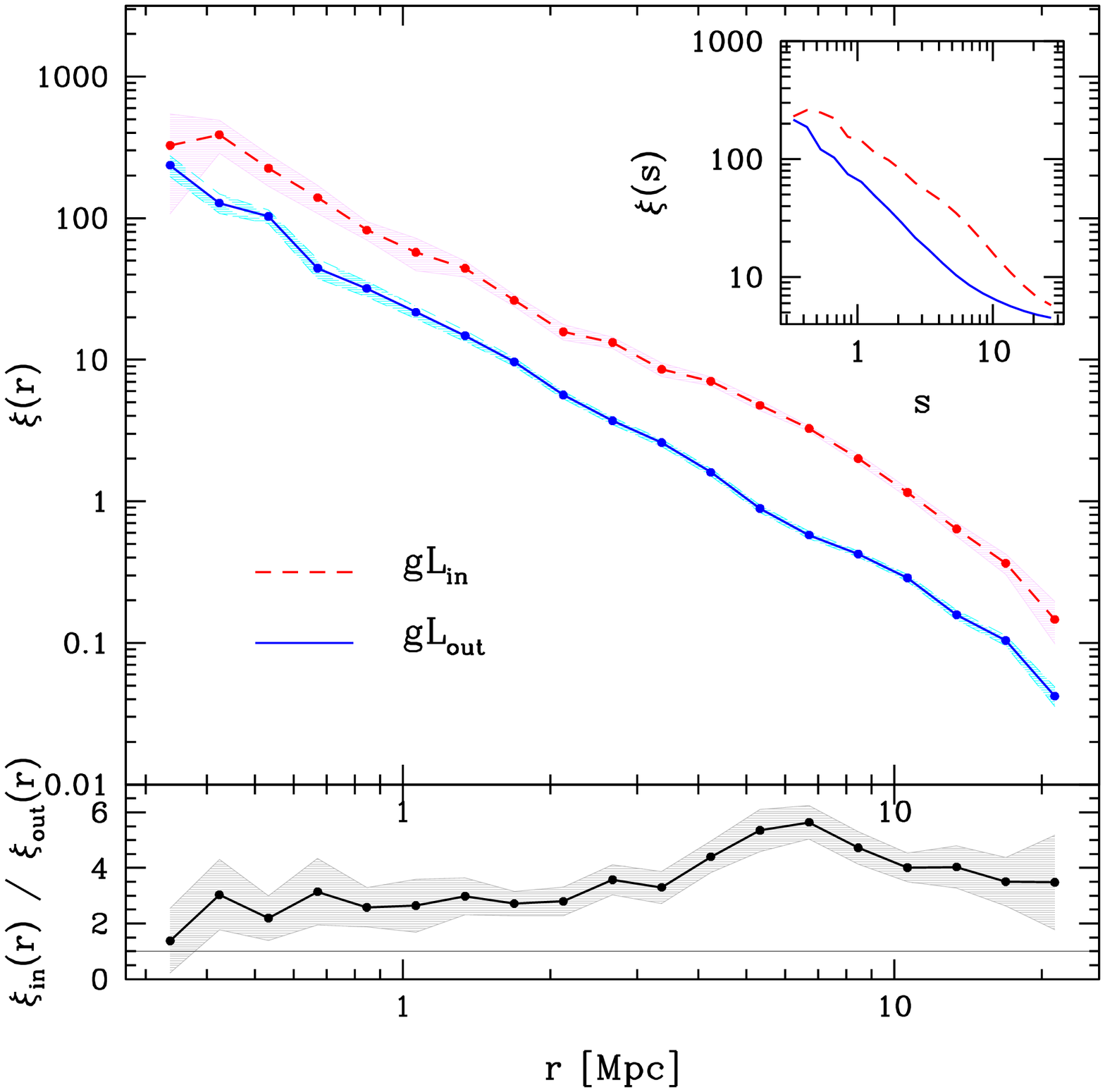}}

   \caption{(a) Luminosity distributions. The left panel corresponds
   to the $g$ samples and the right panel shows the
   luminosity distributions of the new samples ($gL$), defined to
   have the same luminosity distributions.  In both cases, the dashed
   curves correspond to the centres galaxies belonging to
   FVS while the solid curves correspond to the centre
   galaxies outside FVS. The bottom panels show the 
   empirical cumulative luminosity distributions for each case.
   (b) Cross--correlation functions of SDSS--DR7 galaxies for
   $gL$ samples. The dashed lines correspond to galaxies in
   FVS and the solid lines correspond to galaxies outside
   FVS. Both samples have the same luminosity
   distributions.} 

   \label{fig:xi_sample_gL} 
\end{figure*}

Fig. \ref{fig:xi_sample_g} shows the cross--correlation functions for
samples $g_{in}$ and $g_{out}$, i.e., all galaxies in the luminosity
range $-23.0<M_r<-21.0$.
The dashed curves show the results obtained for the sample of galaxies in
FVS, $g_{in}$, while the solid lines show the correlation function
obtained for the sample of galaxies not contained in FVS.
The shadowed regions indicate cosmic variance estimates computed using
Jackknife statistics over the FVS catalogue.
The number of Jackknife realizations is equal to the number of FVS in each case.
As can be seen from this plot, the probability excess of finding a
centre--tracer pair of galaxies is higher for the sample of galaxies
contained in FVS than that corresponding to the sample of galaxies
outside FVS, i.e. the clustering of galaxies is greater when they are
contained in superstructures.
A difference in clustering on large scales (two--halo term) is 
expected given the selection criteria applied to the samples.
Galaxies in the same range of brightness are more clustered if they
are in FVS, on small and large scales (one--halo term and
two-halo term).

From Fig. \ref{fig:xi_sample_g} it is evident that the clustering of galaxies
is greater for galaxies located within the FVS, which suggests the possibility 
that the formation histories of galaxies and their collapse, are strongly 
influenced by the large-scale environment. 
However, this signal should be taken as an upper limit for this effect.  
Indeed, other correlations caused by differences in the galaxy selection of the
two samples could diminish the clustering difference.
In order to quantify how this selection affects the results, we will
consider a number of restrictions on galaxy properties, using other 
samples of Table \ref{table:description}.

%+++++++++++++++++++++++++++++++++++++++++++++++++++++++++++++++++++++++
\subsection{Clustering of faint galaxies around bright galaxies: 
uncovering galaxy luminosity effects} 
\label{sample_gL}

The clustering of galaxies is known to depend on the luminosity of
galaxies \citep{Alimi_1988,
Zehavi:2005, Swanson_2008, Wang:2011, Zehavi:2010, Ross_2011}.
Therefore we will analyse how this dependence affects the previous results.
The median luminosity of $g_{out}$ galaxies is fainter than that of 
$g_{in}$ galaxies. 
This means that centres outside FVS are typically less luminous, and 
therefore their clustering amplitude would be lower than that of 
galaxies which are located in FVS due to this. 
To rule this out, we define samples $gL_{in}$ and $gL_{out}$ from 
the original $g$ samples, trimming objects so that the final luminosity 
distributions are similar. 
This proccess is made by comparing the original luminosity
distributions and then randomly removing objects from either
of these two samples.  This removal is proportional to the
differences between the 
distributions at a given luminosity. We progressively discard objects until the two distributions 
are reasonably similar.

The left panel of Fig. \ref{fig:xi_sample_gL:a} shows the luminosity 
distributions of the original $g$ samples, where dashed curves 
correspond to sample $g_{in}$ and solid curves correspond
to the sample $g_{out}$.
In the right panel we show the luminosity distributions of the
new samples $gL$, trimmed to have the same luminosity distributions,
where dashed curves correspond to $gL_{in}$ 
and solid curves correspond to $gL_{out}$.
The bottom panels show the empirical
cumulative luminosity distributions for each case.

We estimate the cross--correlation functions for both $gL$ samples, 
shown in Fig.\ref{fig:xi_sample_gL:b}.
Here we use the same line style coding of Figure \ref{fig:xi_sample_g}.
As can be seen there is a statistically significant difference between 
the clustering amplitudes of both samples, being greater for 
galaxies contained within FVS.
This signal is observed at all scales including small scales 
(one--halo term), suggesting an effect of the surrounding FVS on 
the galaxy environment.
This signal is, by construction of the samples $gL$, 
independent of the luminosity of centre galaxies but it still 
may depend on the mass of haloes they inhabit.

%+++++++++++++++++++++++++++++++++++++++++++++++++++++++++++++++++++++++
\subsection{Clustering of faint galaxies around bright galaxies: 
Galaxies in groups}
\label{sample_gG}

In this section we study the origin of the previous signal of 
different clustering amplitude for galaxies of equal luminosity 
inside and outside FVS, focusing on the virial mass of the 
host groups of galaxies, which can be interpreted as probing 
larger, but still intermediate scale structures.
We select galaxy groups for which we have estimates of their 
mass.
As a first step, we compare the cross--correlation function of galaxies
in groups, distinguishing between groups that are located inside and
outside FVS.
To this end we use the SDSS--DR7 galaxy group catalogue described in
Section \ref{galaxy_groups}.
We define two new samples of centre galaxies contained in these
groups, one with groups belonging to the FVS ($g_{G_{in}}$) and the other
one with groups that do not belong to any FVS ($g_{G_{out}}$).
Galaxies are again restricted to luminosities within \mbox{$-23.0 < M_{r} <
-21.0$}, as in the previous analyses.
For a more direct comparison, we use the same tracer sample as in
previous sections.

\begin{figure} 
   \centering
   \includegraphics[width=0.5\textwidth]{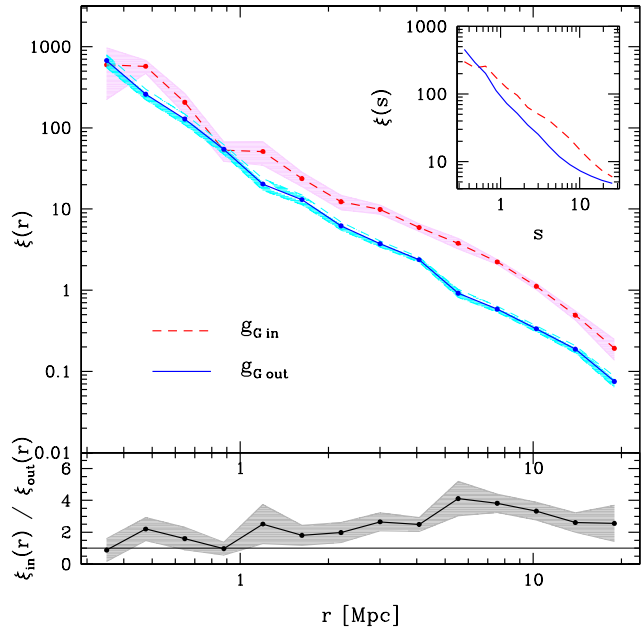}

   \caption{Cross--correlation functions of galaxies in groups 
   ($g_{G}$ samples). The dashed lines correspond to galaxies in 
   $g_{G_{in}}$ sample and the solid lines correspond to galaxies 
   in $g_{G_{out}}$ sample.}
   \label{fig:xi_sample_gG} 
\end{figure}

The cross--correlation functions of the centre--tracer pairs for both
samples are shown in Fig. \ref{fig:xi_sample_gG}.
The dashed curves correspond to sample $g_{G_{in}}$, 
while the solid curves correspond to galaxies in sample
$g_{G_{out}}$.

%+++++++++++++++++++++++++++++++++++++++++++++++++++++++++++++++++++++++
\subsection{Clustering of faint galaxies around bright galaxies: 
uncovering dark matter mass effects}
\label{sample_gGM}

In this section, we analyse the possible effects coming from 
differences in the mass of the groups where the centre galaxies reside.  
Therefore, we repeat the analysis performed for the luminosity dependence
(Section \ref{sample_gL}), redefining instead samples with the same 
mass distributions.
To determine the mass distributions we use the SDSS--DR7 galaxy groups
catalogue described in \cite{Zapata:2009} as in the previous section.
We define two samples of galaxy groups, one with groups belonging to 
FVS and other one with groups outside FVS.
The median of the mass distribution of the galaxy groups that do not
belong to FVS corresponds to a lower value than the median of the mass
distribution of groups inside FVS.
We adjust these samples and redefine two new samples
of groups, trimmed so that their mass distributions are similar.

\begin{figure*}
   \centering
   \subfigure[]{
   \label{fig:xi_C1_galsingroups_sm:a}
   \includegraphics[width=0.48\textwidth]{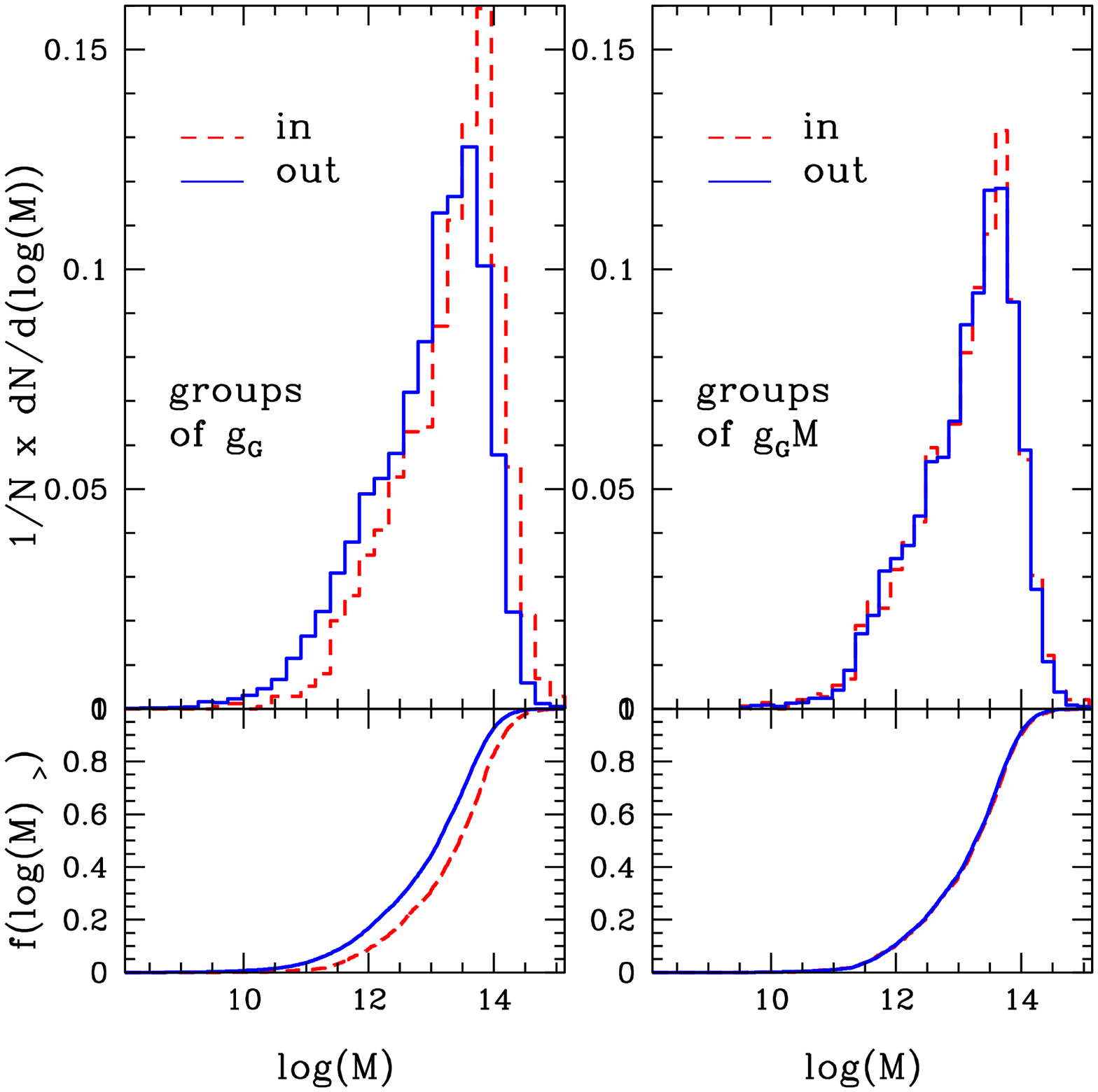}}
   \hspace{5pt}
   \subfigure[]{
   \label{fig:xi_C1_galsingroups_sm:b}
   \includegraphics[width=0.48\textwidth]{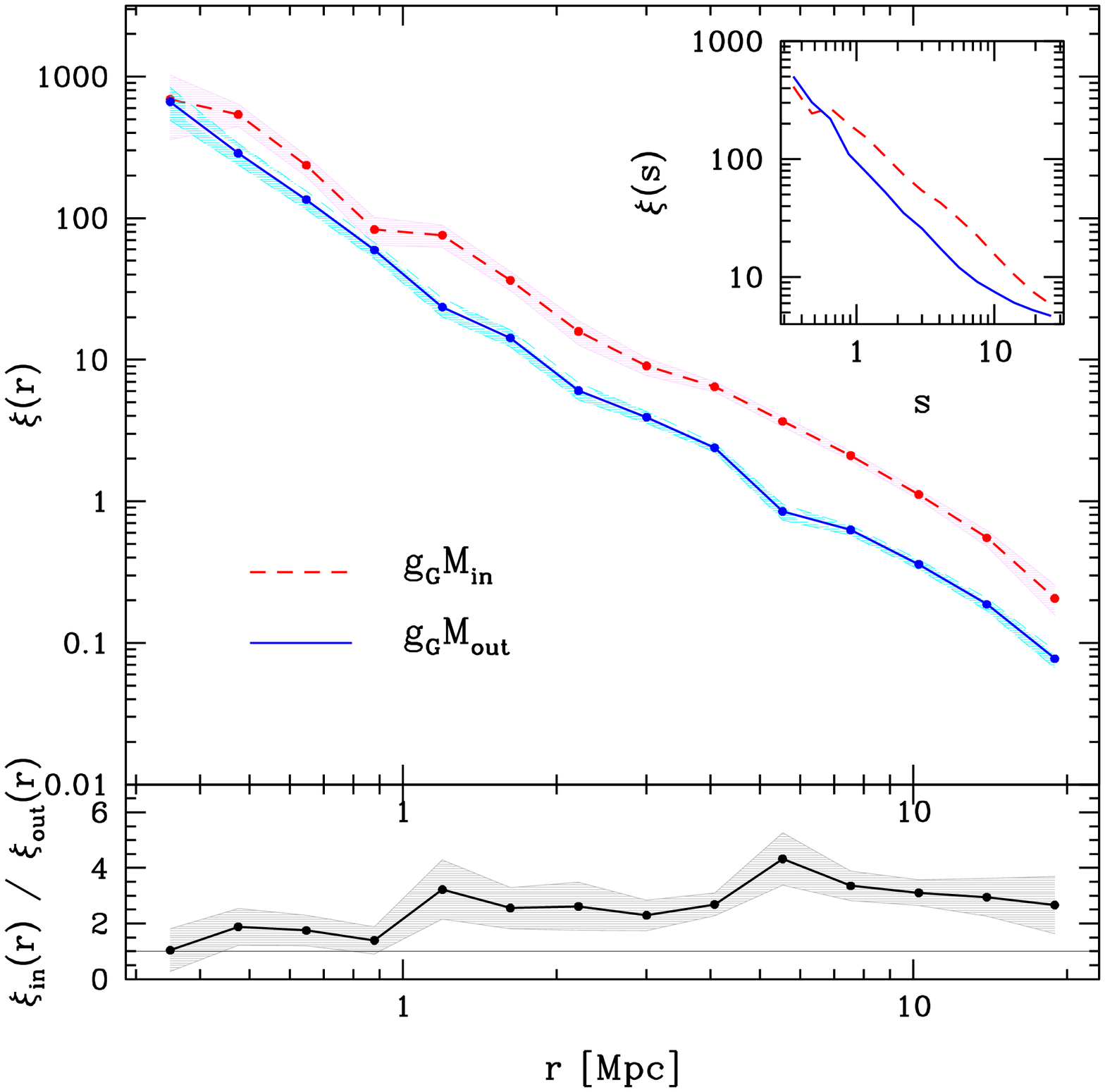}}

   \caption{%
   (a) Mass distributions of galaxy groups.
   The left panel corresponds to the original samples of galaxy
   groups, where the dashed lines are the groups located inside FVS
   and the solid lines are the groups located outside FVS.
   The right panel corresponds to the new samples of groups redefined
   to have the same mass distributions. The bottom panels show the
   empirical cumulative mass distributions for each case.
   (b) Cross--correlation functions of galaxies in groups for $g_{G}M$
   samples.
   The dashed lines correspond to galaxies in groups of $g_{G}M_{in}$ 
   sample and the solid lines correspond to galaxies in groups
   of $g_{G}M_{out}$ sample.
   Both samples have the same mass distributions.}

   \label{fig:xi_C1_galsingroups_sm} 
\end{figure*}

The left panel of Fig. \ref{fig:xi_C1_galsingroups_sm:a} shows the 
mass distributions for the original samples, where dashed lines 
correspond to groups located in FVS and solid lines correspond 
to groups outside FVS.
With the same line style coding, the right panel shows the mass
distributions of the new samples, redefined to have the same mass
distributions.
The bottom panels show the empirical cumulative mass distributions 
for each case.

From the latter two samples of groups, we define two samples of 
centre galaxies contained in these groups:
one with galaxies in groups belonging to FVS ($g_{G}M_{in}$) 
and the other one with galaxies in groups outside FVS ($g_{G}M_{out}$).
Fig. \ref{fig:xi_C1_galsingroups_sm:b} shows the cross--correlation
functions of these samples.
The dashed lines correspond to the sample $g_{G}M_{in}$ while 
the solid lines correspond to the sample $g_{G}M_{out}$.
As can be seen from this figure, the correlation function amplitudes 
differ only slightly in the inner regions \mbox{(r $\lesssim$ 1 Mpc)}.
This is a similar behaviour to that observed in the case of centre
galaxies in groups (\mbox{$-23.0 < M_{r} < -21.0$}), 
selected without restrictions in group mass.

%++++++++++++++++++++++++++++++++++++++++++++++++++++++++++++++++++++++++
\subsection{Clustering of faint galaxies around bright galaxies: 
combined effects from luminosity and mass}
\label{sample_gGML}

In this section we analyse how the luminosity and mass dependence of 
clustering affect the previous results.
To do this, we repeat the previous process restricting samples to have 
comparable galaxy luminosity and host group mass distributions simultaneously.

\begin{figure*}
   \centering
   \subfigure[]{
   \label{fig:xi_C1_galsingroups_sm_sl:a} 
   \includegraphics[width=0.48\textwidth]{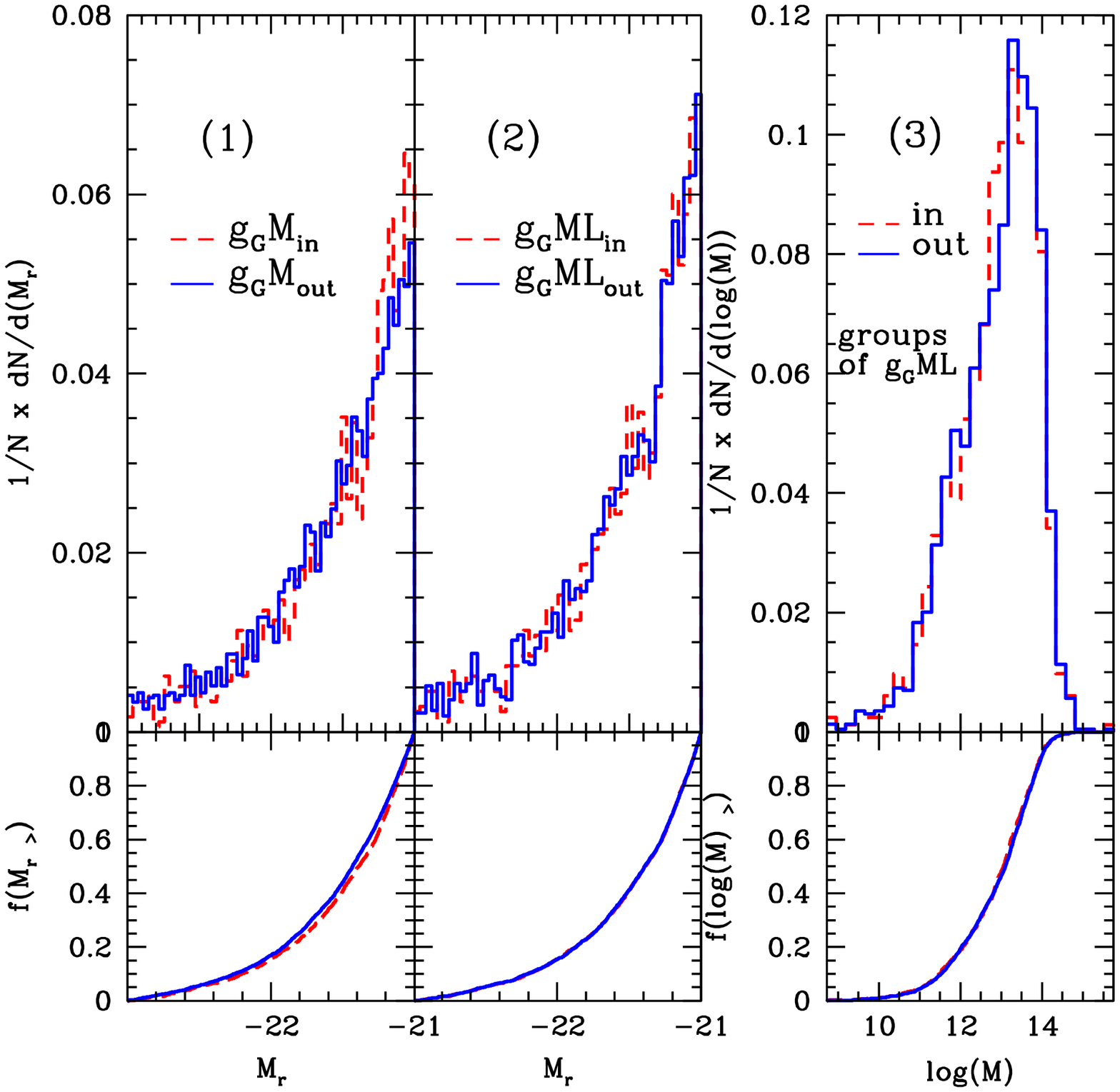}}
   \hspace{5pt}
   \subfigure[]{
     \label{fig:xi_C1_galsingroups_sm_sl:b} 
   \includegraphics[width=0.48\textwidth]{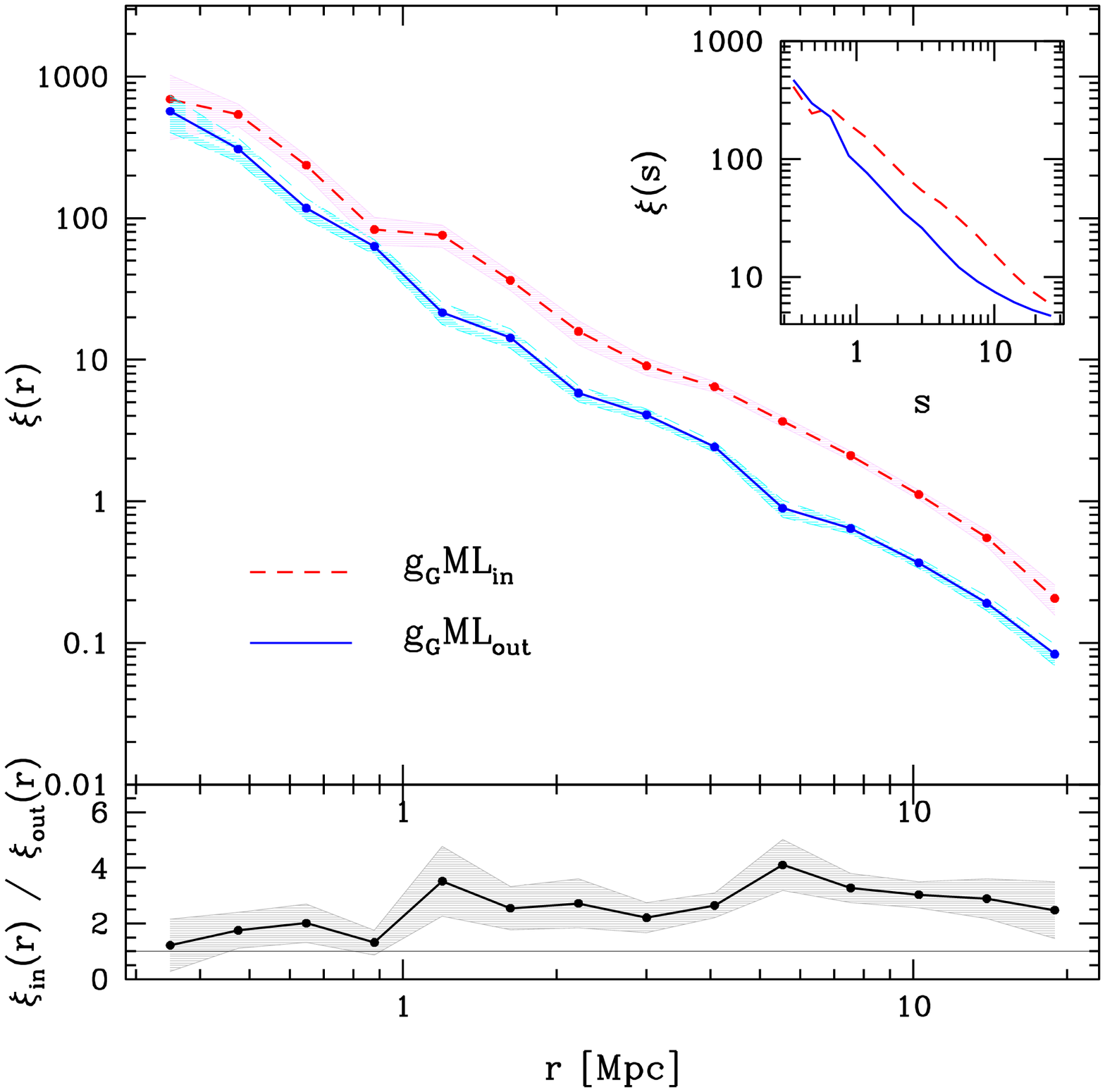}}

   \caption{%
   (a) The left panel shows the luminosity functions of the centre
   galaxies in groups, where these groups have the same mass
   distributions (i.e., luminosity distribution of samples $g_{G}M$). 
   The middle panel shows the luminosity
   distributions of the centre galaxies of the new samples redefined to
   have the same mass and luminosity distributions ($g_{G}ML$). The
   right panel shows the mass distributions for groups composing samples 
   $g_{G}ML$. This plot is to be sure that the mass distributions do not
   change when we adjusted the samples to have the same luminosity
   distributions. The bottom panels show the cumulative functions
   for each case. As the previous cases, in all panels the dashed
   lines indicate groups located in FVS, while the solid 
   lines indicate groups located outside FVS.
  (b) Cross--correlation functions of galaxies in groups for
  $g_{G}ML$ samples. The dashed lines correspond to galaxies in sample 
  $g_{G}ML_{in}$ and the solid lines correspond to
  galaxies in sample $g_{G}ML_{out}$. Both samples
  have the same luminosity and mass distributions.}
\end{figure*}

The left panel of Fig.\ref{fig:xi_C1_galsingroups_sm_sl:a} shows the
luminosity distributions of the centre galaxies of the samples $g_GM$
described in previous section, which are defined to have 
similar mass distributions.
The dashed curves correspond to galaxies in the sample $g_GM_{in}$ while 
the solid curves correspond to galaxies in the sample $g_GM_{out}$.
The middle panel of Fig. \ref{fig:xi_C1_galsingroups_sm_sl:a} 
shows the luminosity distributions of the new samples 
($g_GML_{in}$ and $g_GML_{out}$) defined by equating the luminosity 
distributions of the previous samples (same line coding).
The bottom panels show the empirical cumulative distributions 
for each case.

It is important to ensure that the mass distributions of the 
hosts of centre galaxies remain similar once the luminosity 
distributions are trimmed to be comparable.  
The right panel of Fig.\ref{fig:xi_C1_galsingroups_sm_sl:a} 
shows the mass distributions of the groups of the samples $g_GML$.
The line coding is the same as in the previous plots.
From this plot we can corroborate that the mass distributions of both
samples remain comparable.
The bottom panel shows the empirical cumulative mass distributions
for each case. 
Thus, we obtain two samples of centre galaxies with similar 
luminosity distributions, populating groups with comparable mass 
distributions. 
One of these samples contains galaxies in groups
which are members of FVS ($g_GML_{in}$), and the other comprises  
galaxies in groups away from FVS ($g_GML_{out}$). 
The cross--correlation functions of the centre--tracer pairs for both
samples are shown in Fig.\ref{fig:xi_C1_galsingroups_sm_sl:b}. 
The dashed curves correspond to galaxies in sample $g_GML_{in}$, 
while the solid curves correspond to the galaxies in sample $g_GML_{out}$.

From this figure we conclude that at small scales, the amplitudes of
the clustering of both samples are slightly different, but within a
1-$\sigma$ uncertainty.

%++++++++++++++++++++++++++++++++++++++++++++++++++++++++++++++++++++++++
\subsection{Clustering of faint galaxies around centres of groups: $G8M$ samples. 
} \label{sample_G8M}

In this section we analyse the group--galaxy cross--correlation function, 
considering faint galaxies as tracers, as in the previous cases, but 
in this case selecting geometrical centres of groups instead of 
the brightest group galaxies, i.e, we study the dependence of the 
clustering of faint galaxies around centres of groups, 
using the same faint galaxy tracers as previous sections.

\begin{figure*}
   \centering
   \subfigure[]{
   \label{fig:xi_gal_grps_8m_sm:a} 
   \includegraphics[width=0.48\textwidth]{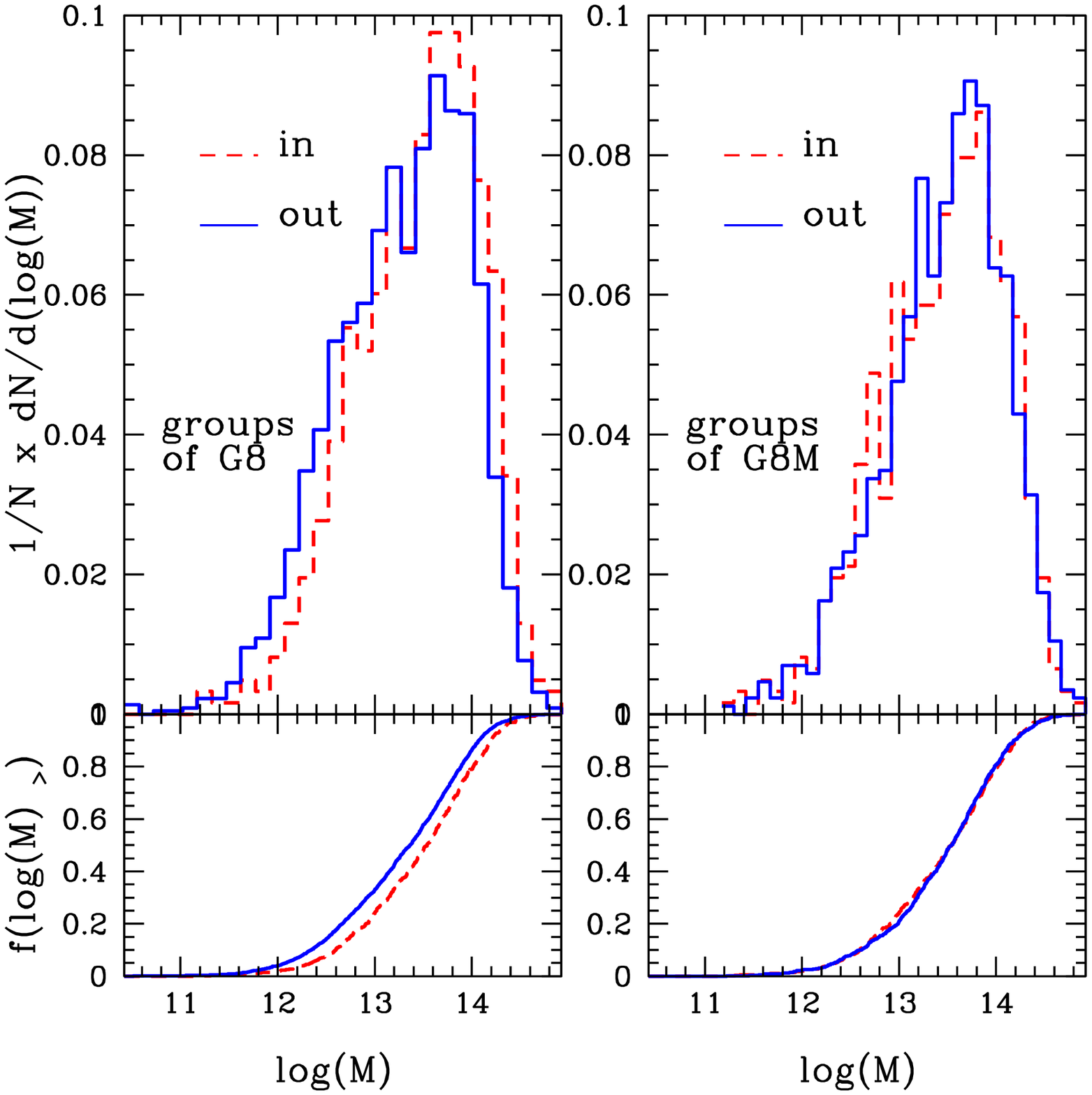}}
   \hspace{5pt}
   \subfigure[]{
   \label{fig:xi_gal_grps_8m_sm:b} 
   \includegraphics[width=0.48\textwidth]{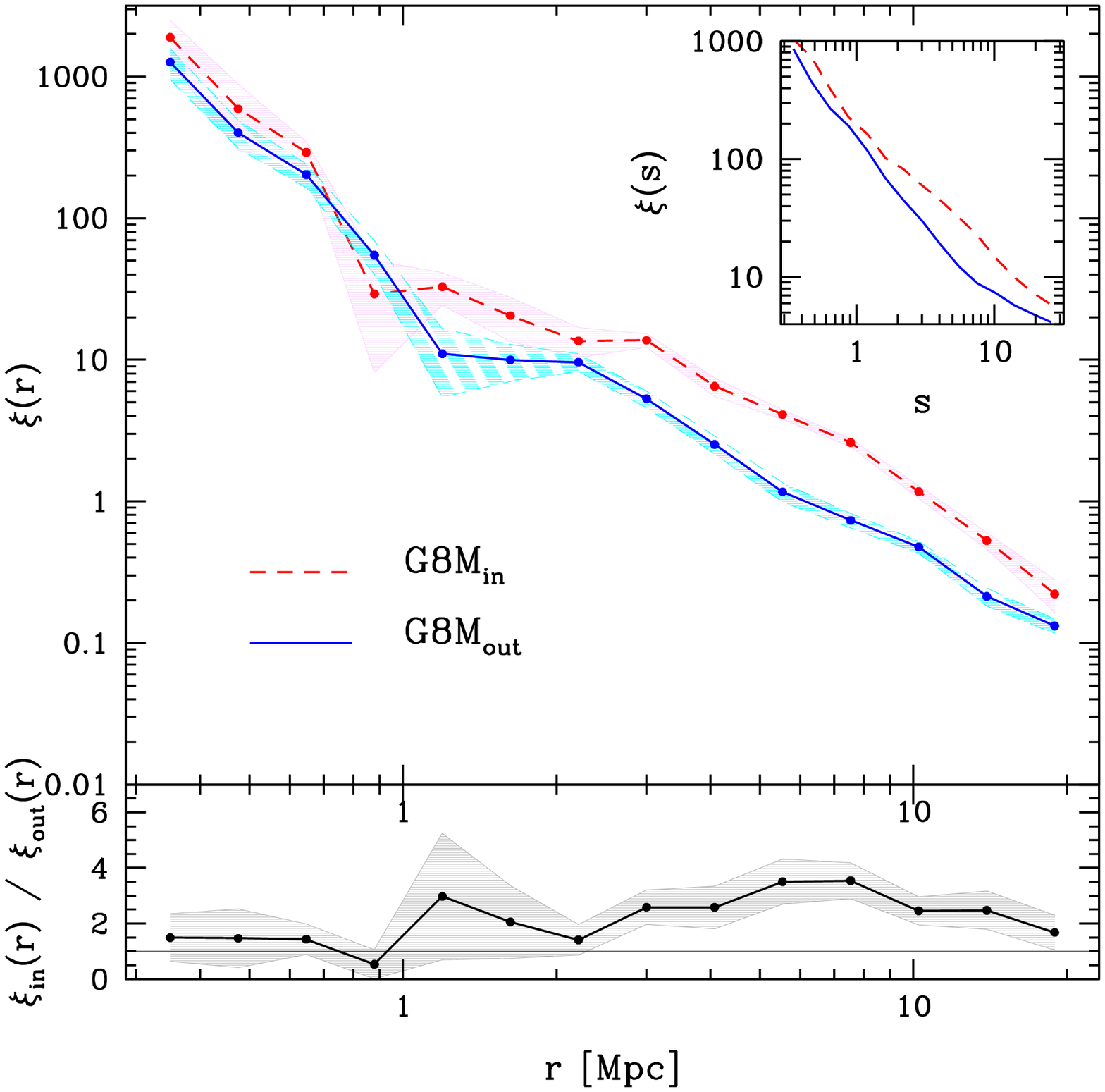}}
   \caption{(a): The left panel shows the mass distributions of groups with at 
least 8 members. The right panel shows the mass distributions of the new samples 
of groups with at least 8 members, adjusted to have the same mass distributions ($G8M$). 
The bottom panels show the empirical cumulative mass distributions for each case. 
In all panels, the dashed lines correspond to groups inside FVS 
and the solid lines correspond to groups outside FVS. 
(b): Group--galaxy cross--correlation functions for samples $G8M$, where the groups 
have at least 8 members and have the same mass distributions. The dashed 
lines correspond to groups in sample $G8M_{in}$ and the solid lines 
correspond to groups in sample $G8M_{out}$. 
  }
\end{figure*}

To this aim, we define two samples of groups with at least 8 galaxy 
members, one with groups located inside FVS and another with 
groups outside FVS.
To make a direct comparison, we define two new samples equating the 
mass distributions of the above mentioned samples.
The left panel of the Fig.\ref{fig:xi_gal_grps_8m_sm:a} shows the mass 
distributions of groups with at least 8 members, where the dashed lines 
correspond to groups that are members of FVS and the solid lines  
correspond to groups that are not members of FVS.
The right panel shows the mass distributions of the new samples, 
redefined to have the same mass distributions. 
The line coding is the same as in the previous sections. 
The bottom panels show the empirical cumulative mass distributions for each case.

To determine the cross--correlation functions we use faint galaxies as tracers 
(defined in Section \ref{sample_g}) and the geometrical centres of groups as 
centres ($G8M$). 
Fig.\ref{fig:xi_gal_grps_8m_sm:b} shows the cross--correlation functions 
for samples $G8M$.
As in previous cases, the dashed lines correspond to centres of groups in sample 
$G8M_{in}$, while the solid lines correspond to centres of groups in sample 
$G8M_{out}$.
As can be seen, there is no differences between the two correlation functions at 
scales of up to \mbox{2 h$^{-1}$ Mpc} when virial masses are considered.

\subsection{Mock: $g_{G}ML-Mock$ and $G8M-Mock$ samples.} \label{mock}

\begin{figure*}
   \centering
   \subfigure[]{
   \label{fig:xi_C1_galsingroups_sm_sl_mock:a} 
   \includegraphics[width=0.48\textwidth]{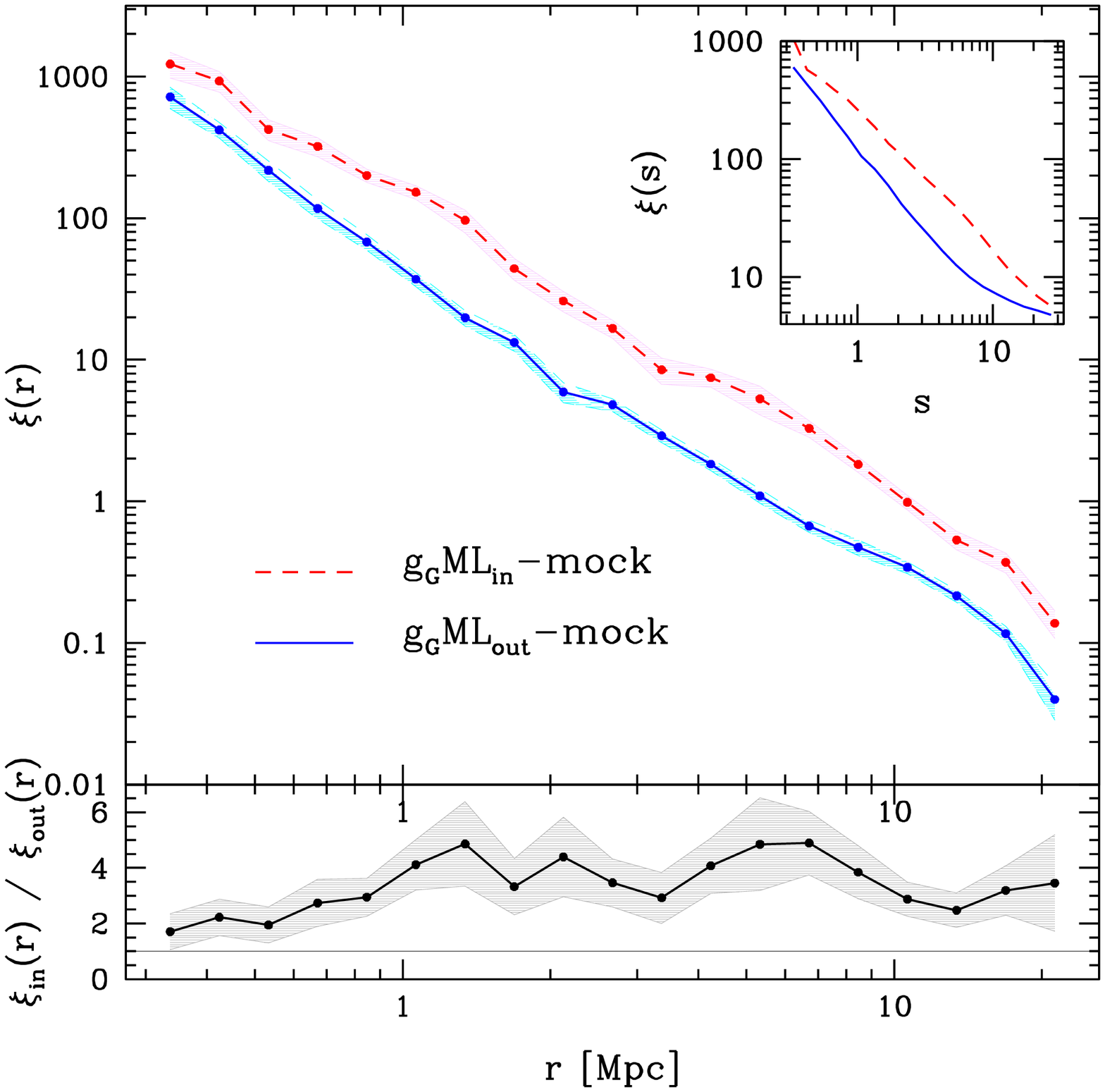}}
   \hspace{5pt}
   \subfigure[]{
   \label{fig:xi_C1_galsingroups_sm_sl_mock:b} 
   \includegraphics[width=0.48\textwidth]{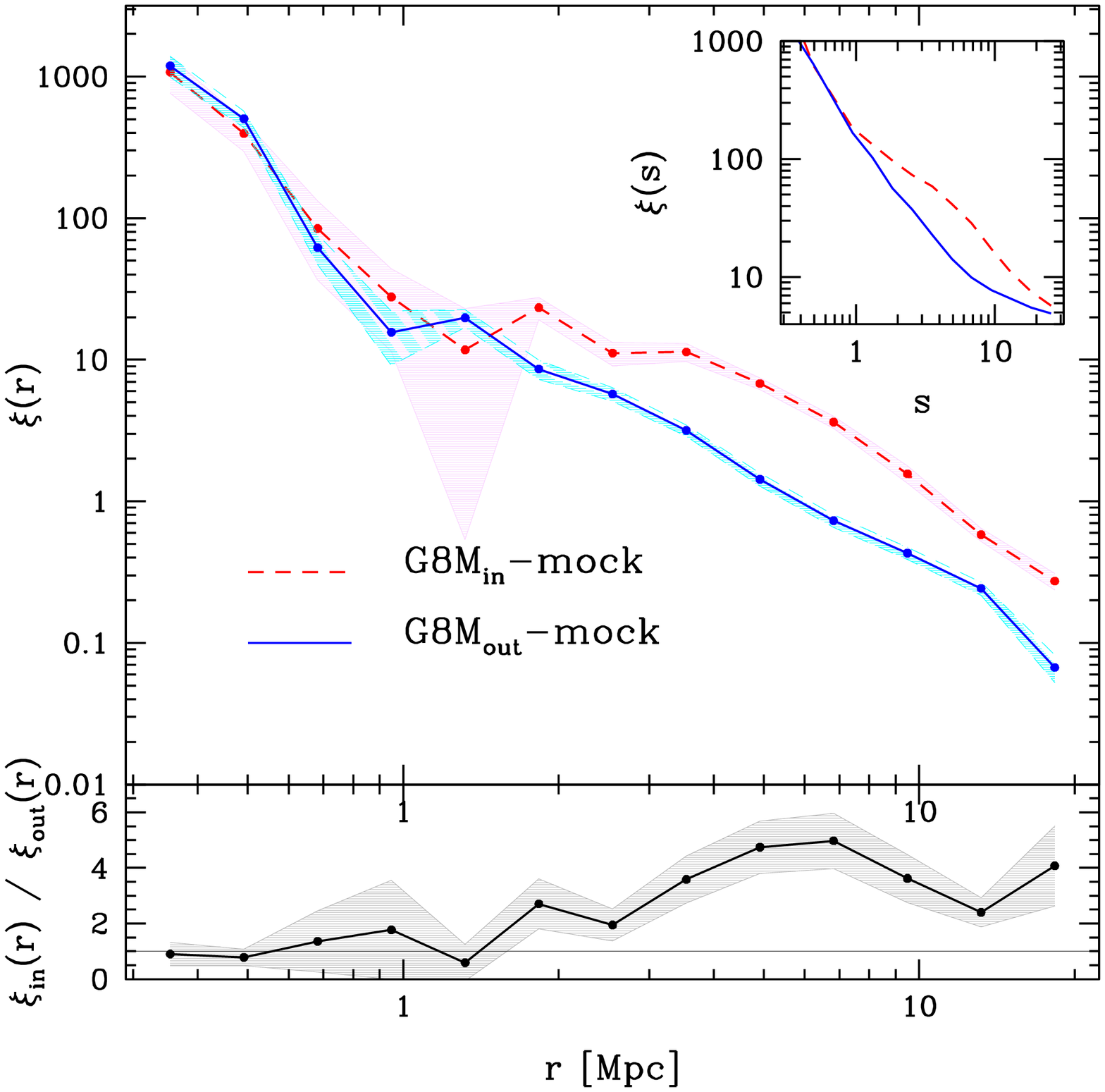}}
   \caption{(a): Mock: Cross--correlation functions of galaxies in groups. The 
   dashed lines correspond to galaxies in sample $g_{G}ML_{in}-mock$ and the 
solid lines correspond to galaxies in sample $g_{G}ML_{out}-mock$. 
Both samples have the same luminosity and mass distributions. 
(b): Mock: Group--galaxy cross--correlation functions, where the groups have at 
least 8 members and have the same mass distributions. The dashed lines correspond 
to groups in sample $G8M_{in}-mock$ and the solid lines correspond to groups in 
sample $G8M_{out}-mock$.}
\end{figure*}%

In order to assess the reproducibility of our previous results by current
models for structure formation, we perform a similar analysis on a mock galaxy
catalogue.   
We use the mock catalogue described in Section \ref{mock_data}, which
implements a semi--analytic model of galaxy formation in a \mbox{$\Lambda$CDM}
cosmological framework. 
We have identified both FVS and groups in this mock galaxy catalogue using the
same procedures than that used in the SDSS--DR7.
We estimate the cross--correlation functions for the corresponding samples to
Section \ref{sample_gGML} and \ref{sample_G8M}. 
The samples \mbox{$g_{G}ML-mock$} are selected from the mock catalogue
following the same steps and taking into account the same restrictions
implemented on the observational data (Section \ref{sample_gGML}). 
Fig. \ref{fig:xi_C1_galsingroups_sm_sl_mock:a} shows the corresponding
cross--correlation functions, using the same line coding as in
previous plots. 
Similarly, the samples \mbox{$G8M-mock$} are defined from the mock catalogue
under the same conditions mentioned in Section \ref{sample_G8M}. 
Fig. \ref{fig:xi_C1_galsingroups_sm_sl_mock:b} shows the cross--correlation
functions obtained from these samples.   
As can be seen in these figures, the correlation functions estimated in mock
catalogues are consistent with the results from observational data.

\subsection{Discussion} \label{bias} 

The mass contained in the FVS is likely to form part of a virialized structure
in the future, but can also be considered to be part of the original
overdensity when it started its collapse in the past. 
Therefore, this mass can be used as a proxy for the equivalent peak height of
the FVS. 
We will test this by comparing the amplitude of the correlation function at
large separations, \mbox{r $>$ 10 h$^{-1}$ Mpc}, with that expected for peaks
corresponding to the FVS masses.

\begin{figure} \centering
\includegraphics[width=0.5\textwidth]{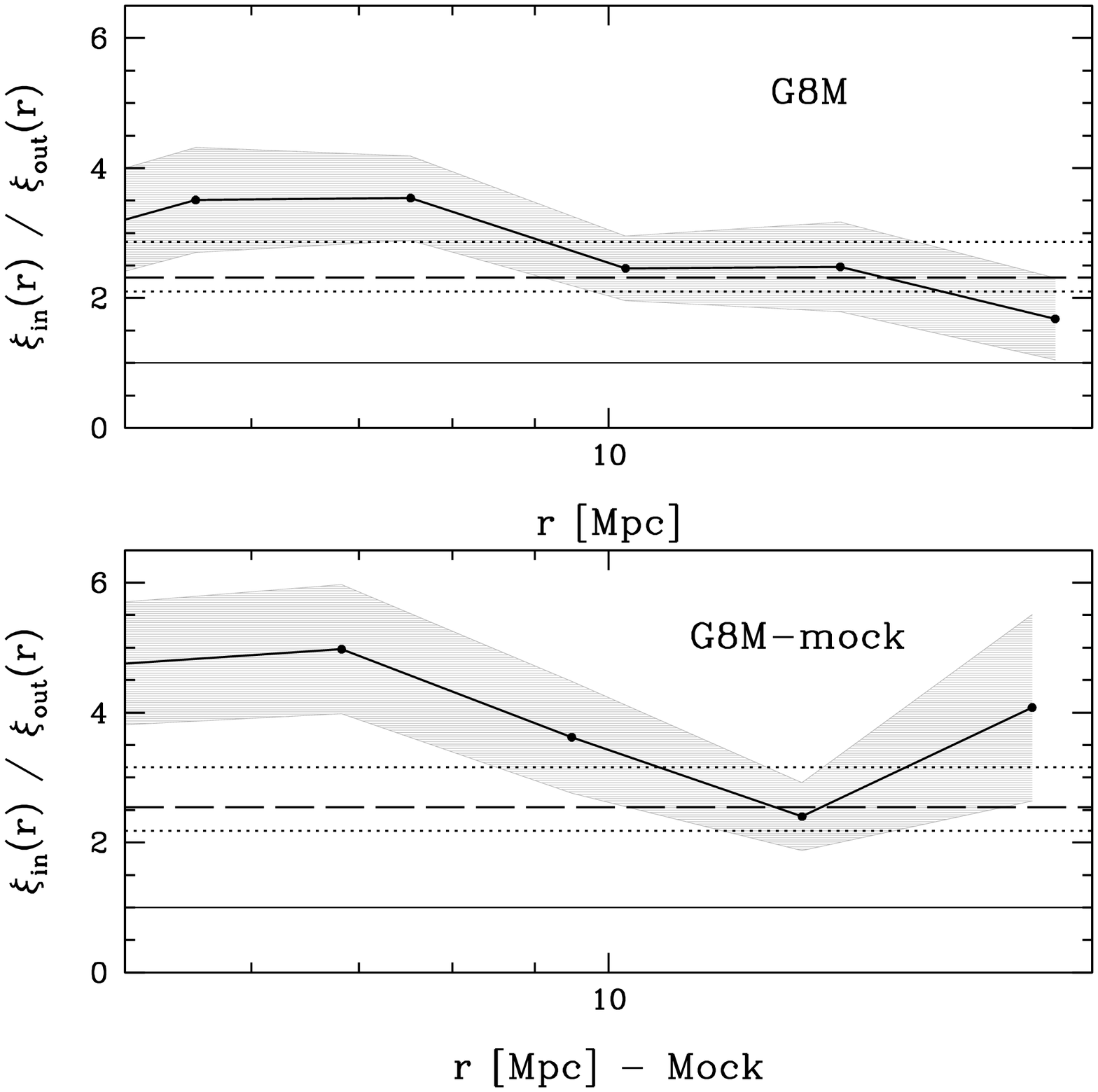} \caption{Ratio
of the real--space correlation functions of galaxies inside and outside FVS,
for samples $G8M$ (upper panel) and $G8M-mock$ (bottom panel), in solid lines,
with uncertainties given by the shaded regions.  The long-dashed horizontal
lines shows the expected value for this ratio using the estimated FVS mass as
that of the original overdensity that will eventually give rise to a virialized
structure, and that of the centres outside FVS.  The dotted lines enclose the
$1-\sigma$ confidence values for this theoretical ratio.} \label{fig:cocientes}
\end{figure}

Fig. \ref{fig:cocientes} shows the ratios between the correlation functions
around objects inside and outside FVS, for $G8M$ and $G8M-mock$ samples. 
These ratios can be compared with the ratio between the corresponding bias
factors expected for objects of the mass of the FVS and that of the centres
outside superstructures.   In order to do this, we need to obtain an
estimate of the groups outside superstructures, and also of the FVSs.
For the former, we use their virial masses, and for the latter,
we measure their total luminosities and adopt the average
mass--to--light ratio for large--scale structures, \mbox {$<M/L> = 577$
M$_{\odot}$/L$_{\odot}$}, given by \cite{Tinker:2010}. 
The bias factors for the groups and FVSs are then obtained using the fitting formula given by
\cite{Tinker:2010}, based on \cite{Sheth:2001}.
The resulting ratio, \mbox{$\sim$ 2.4 $\pm 0.3$}, is shown as a solid
horizontal long-dashed line.  As can be seen, in both real and mock data,
these estimates are consistent with the ratios between the measured correlation
functions.
Errors are represented by the horizontal dotted lines, located at the 16th and
84th percentils of the distribution of bias values for the set of haloes considered.

The amplitude of the correlation function at large separations for objects
lying inside FVS, is consistent with that expected for a structure of the mass
of the FVS, and not with that expected for their virial mass. 
The interpretation for this result is that the positions of the centres used
for the measurement of the correlation function trace, on average, the location
of the overdensity of the FVS.

We test this hipothesis by measuring the centre of mass of the FVS using low
and high mass groups living in them (we define low mass groups as \mbox{$M <
10^{13}$ M$_{\odot}$}, and high mass groups as \mbox{$M > 10^{14}$
M$_{\odot}$}). 
We find that the centres of mass are similar to within \mbox{1.1 h$^{-1}$ Mpc},
which explains the small effect of the centres on the amplitude of the
correlation function at large separations.

\begin{figure} 
   \centering
   \includegraphics[width=0.5\textwidth]{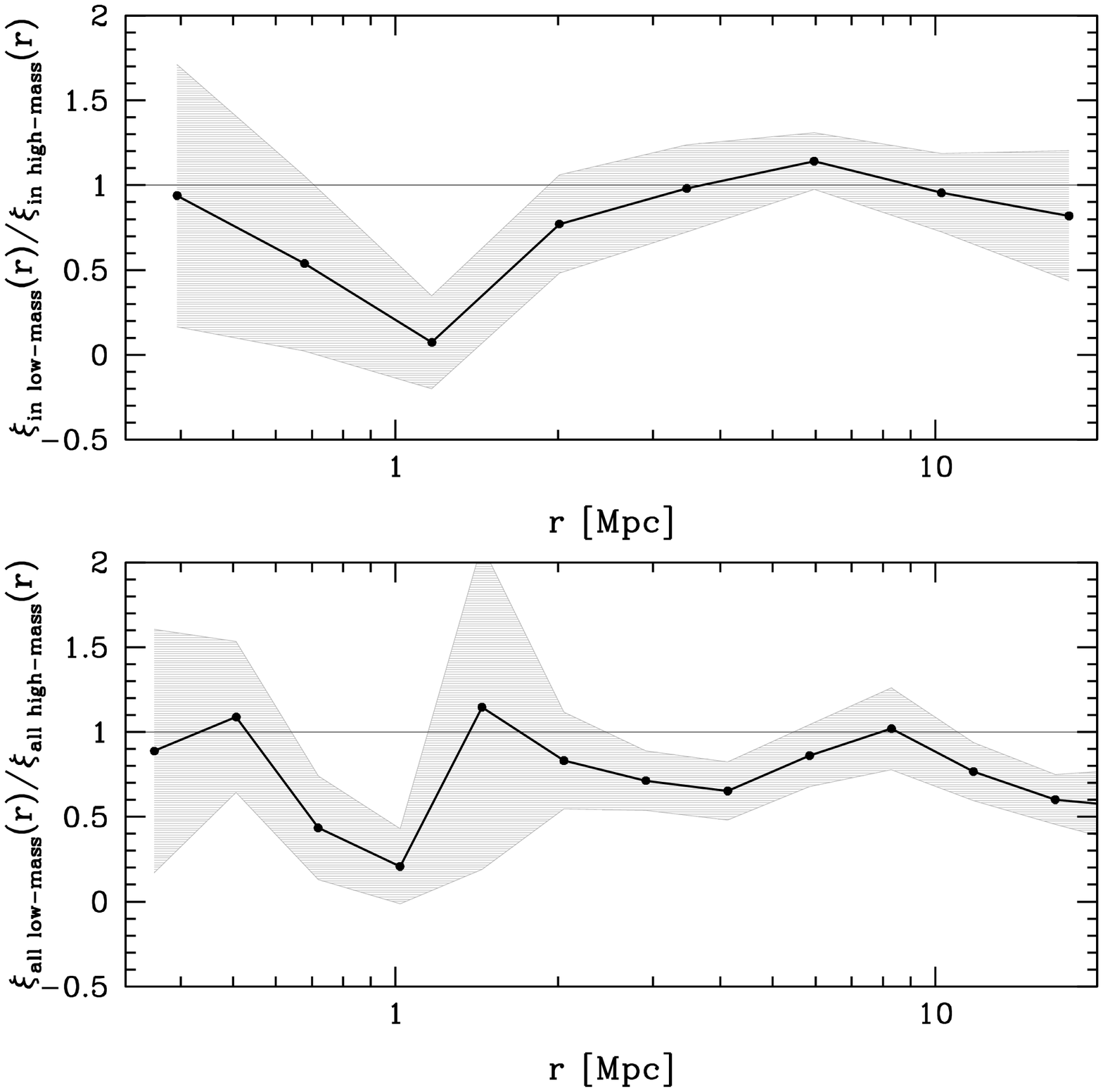}
   \caption{Upper panel: cross-correlation function ratio,
$\xi_{in-low-mass}(r) / \xi_{in-high-mass}(r)$, for groups living inside FVS. 
Low mass corresponds to groups with \mbox{$M < 10^{13}$ M$_{\odot}$} while 
high mass corresponds to groups with \mbox{$M > 10^{14}$ M$_{\odot}$}.
Bottom panel: same as top panel, but for all the groups 
in SDSS separated in low and high mass samples.} 
   \label{fig:cocientes_inout} 
\end{figure}

We also measure the ratio between the correlation functions of these low and
high mass groups living inside FVS.  If they only traced the overdensity of the
FVS, this ratio should be consistent with $1$.
Figure \ref{fig:cocientes_inout} shows this ratio in the top panel, confirming
our hypothesis.  
The lower panel shows the same ratio taking as centres all the low and high
mass groups in the SDSS, and as can be seen, the ratio is significantly lower
than $1$, on average, as expected for this ratio in the case that the groups trace their
own lower or higher overdensity corresponding to their lower or higher masses
(respectively), rather than that of an FVS.

As a final test, we also divided the sample of FVS in two, one for low
and another for high FVS mass samples, and calculated the
cross-correlation function between groups populating these samples and
tracer galaxies. 
As before, we ensure that the mass distributions of the
groups are similar, to avoid the possible influence of the group mass
on the amplitude of the correlation sample (even though we already
showed that this is negligible when groups reside in an FVS). 
We find indications that the clustering for groups in higher mass FVSs is
higher, also lending more support to the results shown previously.

Regarding smaller scales, those corresponding to the one--halo term,
our analysis
indicates that the differences in the clustering amplitude, taking into
account whether galaxies reside in FVS or not, are the sum of several
contributions.
We identified at least two sources that may be responsible for these
differences, i.e., the different clustering amplitude due to galaxy
luminosity and galaxy host mass selection effects.
We acknowledge that there may be other sources that contribute to this
deviation related to the selection of distinct large--scale
environments.  
However, our analysis indicates that deviations originated on those
sources are not statistically significant, pointing at a scenario where the
FVS environnment has no measurable effect on the 1-halo clustering.

%}}}S_results/*

\section{Conclusions} \label{s_conclusus} 
%{{{S_conclusus/*

In this work we present a detailed analysis of cross--correlation of faint and
bright galaxies and its dependence on large--scale structures. 
To this end, we use the properties of galaxy groups to characterize local
environments \citep{Zapata:2009}, and the properties of future virialized
structures \citep{Luparello:2011} as proxies of large scale environments.
We distinguish between galaxies belonging to FVS and galaxies outside FVS in a
volume limited sample up to $z=0.12$.
At large scales, the amplitude of the two point cross--correlation function of
bright and faint galaxies is significantly larger when the bright galaxies
reside in FVS. 
At small scales, where the clustering cross--correlation signal is dominated by
the local environment, the results inside and outside FVS are slightly different.
In order to disentangle local and large scale contributions to the observed
clustering, we have considered subsamples where the galaxy luminosity, and host
group mass distributions are forced to be similar.
Once these restrictions have been implemented, the resulting
correlation functions show statistically negligible differences.

In order to asses the reproducibility of these results within current models 
for structure formation, we have performed a similar analysis using a 
semi--analytic implementation in a \mbox{$\Lambda$CDM} cosmological model.
We have determined bright--faint galaxy cross--correlations taking into 
account the local and global environment using groups and FVS in a similar 
fashion as performed in the observations. 
By considering subsamples with similar galaxy luminosity and host group 
mass distributions we determine that the cross--correlations dependence on  
large-scale structures is consistent with the dependence found in the observations.

This analysis suggests that the resulting behaviour of the cross--correlation
clustering of bright and faint galaxies on large scale structures is a generic
feature of galaxy clustering in hierarchical scenarios and our current
understanding of galaxy formation scenario.

We studied the amplitude of the correlation function at large separation for
centres inside and outside FVS, comparing them to that expected from the
theory.  
We find that objects in FVS show a large--scale clustering consistent with that
of an overdensity of the mass of the FVS in which they reside. 
These objects trace, on average, the centre of mass of the FVS, explaining this
effect. 
Furthermore, for groups inside FVS, the amplitude of the correlation function
at two--halo term separations does not seem to depend on group mass. 
This dependence is more clear when considering all the groups in the SDSS.

In a forthcoming paper we will assess the properties of galaxies in systems
residing inside and outside FVS.

%!///  ///  ///  ///  ///  ///  ///  ///  ///  /// 

%}}}S_conclusus/*
 
%··········································································
%····································································

\section*{acknowledgements}
%{{{/*
This work was partially supported by the
Consejo Nacional de Investigaciones Cient\'{\i}ficas y T\'ecnicas
(CONICET), and the Secretar\'{\i}a de Ciencia y Tecnolog\'{\i}a,
Universidad Nacional de C\'ordoba, Argentina. NP was supproted by
Fondecyt Regular \#1110328.
Funding for the SDSS and SDSS--II has been provided by the Alfred P.
Sloan Foundation, the Participating Institutions, the National Science
Foundation, the U.S. Department of Energy, the National Aeronautics
and Space Administration, the Japanese Monbukagakusho, the Max Planck
Society, and the Higher Education Funding Council for England. The
SDSS Web Site is http://www.sdss.org/.
The SDSS is managed by the Astrophysical Research Consortium for the
Participating Institutions. The Participating Institutions are the 
American Museum of Natural History,
Astrophysical Institute Potsdam, University of Basel, University of
Cambridge, Case Western Reserve University, University of Chicago,
Drexel University, Fermilab, the Institute for Advanced Study, the
Japan Participation Group, Johns Hopkins University, the Joint
Institute for Nuclear Astrophysics, the Kavli Institute for Particle
Astrophysics and Cosmology, the Korean Scientist Group, the Chinese
Academy of Sciences (LAMOST), Los Alamos National Laboratory, the
Max-Planck-Institute for Astronomy (MPIA), the Max-Planck-Institute
for Astrophysics (MPA), New Mexico State University, Ohio State
University, University of Pittsburgh, University of Portsmouth,
Princeton University, the United States Naval Observatory, and the
University of Washington.   
The mock catalogue and related simulations were performed using 
the Geryon cluster at the Centro de Astro-Ingenier\'{\i}a UC.
The Millenium Run simulation used in this paper was carried out by the
Virgo Supercomputing Consortium at the Computer Centre of the
Max--Planck Society in Garching. 
%}}}/*

%\bibliographystyle{mn2e}
%\bibliography{Biblio}

\end{document}